\documentclass[preprint,preprintnumbers,amsmath,amssymb]{revtex4}
\usepackage[dvips]{graphicx}
\usepackage{graphicx}
\usepackage{pstcol}
\usepackage{amsfonts}
\usepackage{bm}
\usepackage{amsmath}
\usepackage{amssymb}
\usepackage{color}
\usepackage[all]{xy}

\def\be{\begin{equation}}
\def\ee{\end{equation}}
\def\bea{\begin{eqnarray}}
\def\eea{\end{eqnarray}}

\begin{document}

\title{Non-extensive and quasi-homogeneous geometrothermodynamics}

\author{Hernando Quevedo}
\email{quevedo@nucleares.unam.mx}
\affiliation{Instituto de Ciencias Nucleares, Universidad Nacional Aut\'onoma de M\'exico, Mexico}
\affiliation{Dipartimento di Fisica and Icra, Universit\`a di Roma “La Sapienza”, Roma, Italy}
\affiliation{Al-Farabi Kazakh National University, Almaty, Kazakhstan}

\author{Mar\'ia N. Quevedo}
\email{maria.quevedo@unimilitar.edu.co} 
\affiliation{Departamento
de Matem\'aticas, Facultad de Ciencias B\'asicas, Universidad
Militar Nueva Granada, Cra 11 No. 101-80, Bogot\'a D.C., Colombia}

\date{\today}

\begin{abstract}

We study the thermodynamic properties of black holes, taking into account the non-extensive character of their entropy at the thermodynamic and statistical level. 
To this end, we assume that the R\'enyi entropy determines the fundamental thermodynamic equation of black holes and is represented by a quasi-homogeneous function. As a consequence, the R\'enyi parameter turns out to be an independent thermodynamic variable, which must be treated in the framework of extended thermodynamics. As a particular example, we use the formalism of geometrothermodynamics to show that the Schwarzschild black hole can become stable for certain values of the R\'enyi parameter.


\end{abstract}


\maketitle

\section{Introduction}
\label{sec:int}

The Bekenstein-Hawking entropy formula is considered as the starting point for the development of black hole thermodynamics 
\cite{bekenstein1972black,hawking1975particle}. This formula establishes that the entropy of a black hole is completely determined by the area of the event horizon, which contrasts with ordinary systems, where the entropy is proportional to the volume of the system. As a consequence, the entropy turns out to be a 
non-extensive quantity.

On the other hand, one of the main results that corroborate the validity of black hole thermodynamics is the discovery by Hawking that, in the semiclassical theory of gravity, a black hole radiates as a black body, and the corresponding temperature coincides with the thermodynamic temperature obtained from the entropy-area relationship. 
This result generates certain skepticism, because to derive Hawking's temperature, the validity of the Boltzmann-Gibbs statistics is assumed, which, however, implies the extensivity of the entropy. 
An additional inconsistency of black hole thermodynamics consists in the assumption that to explore the properties of 
non-extensive black holes, we apply the standard approach of classical extensive thermodynamics. 
An important consequence of this approach is that a static black hole in Einstein gravity is thermodynamically unstable \cite{davies1978thermodynamics}.

The inconsistencies of black hole thermodynamics have been addressed from different points of view, including the Poincar\'e turning point method \cite{arcioni2005stability}, the quantum corrected entropy \cite{czinner2015black}, the non-extensive Tsallis and  R\'enyi entropies in the Schwarzschild spacetime \cite{biro2013q,czinner2016renyi} and in the Schwarzschild-anti-de Sitter spacetime \cite{anusonthi2025thermodynamic}. The conditions for consistently applying generalized non-extensive entropies in black hole thermodynamics have been extensively analyzed in \cite{nojiri2021area,nojiri2022nonextensive,elizalde2025black}. Furthermore, the extensivity of black hole thermodynamics within the AdS/CFT framework
has been studied in \cite{karch2015holographic,mancilla2024generalized}.

On the other hand, in \cite{czinner2016renyi},
it has been shown that at fixed temperature the Schwarzschild black hole can be in stable equilibrium with the surrounding thermal radiation, 
and in \cite{anusonthi2025thermodynamic}, the stability has been shown assuming the cosmological constant and the non-extensive parameter as thermodynamic variables.  
In this work, we will follow the last approach, in which the manifestly non-extensive R\'enyi entropy is used instead of the Bekenstein-Hawking entropy. We will show that to generate stability phases in black holes, it is sufficient to consider the non-extensive parameter as thermodynamic variable using the formalism of geometrothermodynamics. 

This work is organized as follows. In Sec. \ref{sec:non}, we argue that to avoid inconsistencies, it is convenient to adopt the non-extensive R\'enyi entropy in black hole thermodynamics. Moreover, we show that the quasi-homogeneity property of black holes implies that the non-extensivity parameter should be considered as an independent thermodynamic variable in the framework of extended thermodynamics. In Sec. \ref{sec:gtd}, we review the formalism of geometrothermodynamics (GTD), which consistently applies quasi-homogeneity and Legendre invariance to investigate the properties of systems in extended thermodynamics.   
In Sec. \ref{sec:sch}, we apply GTD to analyze the extended Schwarzschild black hole, whose thermodynamic properties are investigated in Sec. \ref{sec:res}, where we show the appearance of a new phase with stable states. Finally, in Sec. \ref{sec:con}, we discuss our results {
and comment on the physical and geometric motivations for choosing the  nonextensivity parameter as a thermodynamic variable. 
}


\section{Non-extensivity and quasi-homogeneity}
\label{sec:non}

The non-extensivity of the Bekenstein-Hawking entropy, which relates the entropy $S_{BH}$ with the area $A$ of the black hole event horizon, $S_{BH} = \frac{A}{4}$, became evident in the very early formulation of black hole thermodynamics \cite{bekenstein1973black,bardeen1973four,hawking1975particle,hawking1976black}, which states that the standard procedures of ordinary classical thermodynamics can be applied to exploit the properties of black holes. 
The starting point to develop the formalism of thermodynamics is to consider the Bekenstein-Hawking entropy as the fundamental equation of the system \cite{callen1991thermodynamics}, i.e., $S_{BH} = S_{BH}(E^a)$ is a function from which all the thermodynamic properties of the system can be derived and, in particular, satisfies the 
first law of thermodynamics
\be
dS_{BH} = I_a  dE^a, \quad
I_a = \frac{\partial S_{BH}}{\partial E^a}.
\ee
Here, $E^a$,  $a=1,2,...,n$, represent the physical parameters that enter the radius of the event horizon, such as mass, angular momentum, etc. The integer $n$ is interpreted as the number of thermodynamic degrees of freedom of the black hole.
The variables $I_a$ are dual to $E^a$ and in ordinary classical thermodynamics are interpreted as intensive variables. Strictly speaking, in black hole thermodynamics, this identification is no longer valid due to the fact that the Bekenstein-Hawking entropy is not a homogeneous function of its variables.

Extensivity of the entropy $S_{BH}(E^a)$ implies that it must satisfy the homogeneity condition $S_{BH} (\lambda E^a) = \lambda S_{BH}(E^a)$, where $\lambda$ is a positive definite constant, representing the scale factor that acts equally on all independent variables $E^a$. In the case of general relativity, the most general black hole in Einstein-Maxwell theory is described by the Kerr-Newman spacetime, in which the radius of the outer event horizon can be expressed as
\be
r_h = M + \frac{1}{M}\sqrt{M^4- J^2 - M^2Q^2 }\ ,
\ee
where $M$ is the mass, $J$ the angular momentum, and $Q$ the electric charge of the black hole. The corresponding Bekenstein-Hawking entropy is given as
\be
S_{BH}= 2M^2-Q^2 + 2\sqrt{M^4-J^2-M^2 Q^2} ,
\label{knent}
\ee
so that in this case $E^a=(M,J,Q)$. The corresponding first law in the entropy representation reads \cite{bardeen1973four}
\be
dS_{BH} = \frac{1}{T}dM - \frac{\Omega}{T}dJ - \frac{\phi}{T}dQ \ ,
\ee
where $\phi$ is the electric potential and $\Omega$ the angular velocity on the horizon.

Rescaling all the variables as $E^a\rightarrow \lambda E^a$ of the entropy (\ref{knent}), one can see that the homogeneity condition is not satisfied, a result that can be associated with the fact that the Bekenstein-Hawking entropy is proportional to area and not to volume as in ordinary thermodynamic systems. Introducing new variables as  $M^2\rightarrow m,\ Q^2\rightarrow q,\ J\rightarrow j$, it is possible to recover homogeneity as $S_{BH}(\lambda m, \lambda j , \lambda q) = \lambda S_{BH}(m,j,q)$. However, such a redefinition of variables is not allowed because it can drastically change the thermodynamic properties of the system \cite{quevedo2019quasi}. It follows that black holes cannot be treated as ordinary homogeneous systems. 

To solve the above inconsistencies using the formalism of geometrothermodynamics, we have proposed in \cite{quevedo2019quasi} to consider, in general, black holes as quasi-homogeneous systems, as originally proposed by Belgiorno in \cite{belgiorno2003black,belgiorno2003quasi}. These are systems that rescale differently for each independent thermodynamic variable and satisfy the relationship 
\be
S_{BH}(\lambda^{\beta_a} E^a ) = \lambda^{\beta_{S_{_{_{BH}}}} }S_{BH} (E^a), 
\label{quasi}
\ee
where the real constants $\beta_a$ are called the coefficients of quasi-homogeneity. In the limiting case $\beta_a=1$, we obtain the usual homogeneous condition. 

Considering the entropy for the Kerr-Newman black hole (\ref{knent}), it is easy to see that the 
quasi-homogeneity condition (\ref{quasi}) is satisfied if the coefficients $\beta_a=(\beta_M,\beta_J,\beta_{S_{BH}})$ are related by
\be
\beta_Q = \beta_M, \quad\beta_J = 2\beta_M , \quad 
\beta_{S_{BH}}= 2\beta_M .
\ee
As a result of the quasi-homogeneity property, the Euler identity turns out to depend explicitly on the coefficients $\beta_a$ as \cite{quevedo2019quasi} 
\be
\beta_M \frac{\partial S_{BH}}{\partial M } M +
\beta_J \frac{\partial S_{BH}}{\partial J} J +
\beta_Q \frac{\partial S_{BH}}{\partial Q} Q  = \beta_{S_{BH}} S_{BH} ,
\ee
which is then equivalent to the Smarr relation
\be
M = 2TS_{BH}+2\Omega J + \phi Q .
\ee


As mentioned above, the expression for the thermodynamic temperature $T=\left(\frac{\partial S_{BH}}{\partial M}\right)^{-1}$, obtained from the non-extensive entropy $S_{BH}$, coincides with the statistical Hawking temperature $T_H$, obtained in semiclassical gravity using the Boltzmann-Gibbs statistics, which is used for extensive systems. 
Several approaches have been proposed to mitigate this problem. In particular, the entropy $S_{BH}$ can be identified with the Tsallis entropy $S_T$, which by definition is non-extensive, also at the statistical level. However, Tsallis entropy is known to be in contradiction with the 0-th law of black hole thermodynamics. However, it has been shown that this problem can be solved using the R\'enyi entropy 
\be
S_R = \frac{1}{r}\ln (1 + r  S_{BH}) \ ,
\label{ren}
\ee
where $r$ is a parameter that can be interpreted as a measure of the deviation of the system from an extensive system. This parameter is defined in the interval $r  \in (-\infty, 1]$ and, in particular, in the limit $r \rightarrow 0$, we recover the Bekenstein-Hawking entropy. 
The main point now is that R\'enyi entropy is non-extensive from both the thermodynamic and statistical points of view.  

According to Eq.(\ref{knent}), the R\'enyi entropy for a Kerr-Newman black hole is given by
\be
S_R = \frac{1}{r}\ln\left[1 + \pi r\left(2M^2-Q^2 + 2\sqrt{M^4-J^2-M^2 Q^2}\right)\right].
\ee
To be in agreement with the above argumentation of black hole thermodynamics, we demand that the R\'enyi entropy be quasi-homogeneous, i.e., $S_R(\lambda^{\beta_a}E^a) = \lambda^{\beta_{_R}} S_R(E^a)$. It is easy to see that this condition implies that the non-extensive parameter $r$ is a thermodynamic variable, i.e., $E^a=(M,J,Q,r)$, and the quasi-homogeneity parameters are related by
\be
\beta_Q = \beta_M, \quad\beta_J = 2\beta_M , \quad 
\beta_r=-2\beta_M, \quad
\beta_{S_{R}}= 2\beta_M .
\ee
In the case of black holes with cosmological constant, the quasi-homogeneity condition also implies that the cosmological constant is a thermodynamic variable related to pressure \cite{quevedo2019quasi}, leading to a completely different phase transition structure of black holes in Einstein gravity \cite{kubizvnak2017black}.
In the following sections, we will explore the consequences of considering  the R\'enyi parameter $r$ as a thermodynamic variable.


\section{Quasi-homogeneous geometrothermodynamics }
\label{sec:gtd}

The formalism of GTD aims to describe the properties of thermodynamic systems using geometric concepts \cite{quevedo2007geometrothermodynamics}. To this end, any thermodynamic system is endowed with an equilibrium space ${\cal E}$, whose points represent the equilibrium states of the system. Moreover, the space ${\cal E}$ is assumed to be differentiable so that it can be equipped with a Riemannian metric $g$. The pair $({\cal E},g)$ constitutes a Riemannian manifold whose geometric properties should be related to the thermodynamic properties of the system. For instance, geodesics of ${\cal E}$ are related to quasistatic processes  \cite{quevedo2015relativistic} and curvature singularities are interpreted as phase transitions \cite{alvarez2008unified}.

Another important ingredient of GTD is that it is invariant with respect to Legendre transformations. This is an important symmetry property of classical thermodynamics, which implies that the properties of a system do not depend on the choice of thermodynamic potential. In fact, different thermodynamic potentials in classical thermodynamics are related by Legendre transformations \cite{callen1991thermodynamics}.  In GTD, we implement Legendre invariance by introducing an additional auxiliary $(2n+1)-$dimensional space ${\cal T}$ called the phase space, where Legendre transformations are represented as diffeomorphisms. The phase space can be equipped with a Riemannian metric $G$, which is demanded to be invariant with respect to Legendre transformations. The details of this approach involve several aspects of contact geometry that can be consulted in \cite{quevedo2007geometrothermodynamics,quevedo2023unified}. As a general result, we observe that there are three classes of metrics $G^I, \ G^{II},$ and $G^{III}$ that are invariant with respect to Legendre transformations.
In turn, the equilibrium space $({\cal E}, g)$ is introduced as an $n-$dimensional subspace of the phase space $({\cal T},g)$ that inherits the property of Legendre invariance. 

To be more specific, let us denote by $E^a$ the coordinates of the equilibrium space. We denote by $\Phi(E^a)$ any thermodynamic potential that can be obtained in the entropy or energy representations by means of Legendre transformations. This means that the set $E^a$ can contain any combination of $n$ independent thermodynamic variables and their duals. In this sense, the results of GTD are independent of the choice of thermodynamic potential and their independent variables. Moreover, any arbitrary thermodynamic potential can be considered as the fundamental equation, $\Phi =\Phi(E^a)$, that describes the corresponding system and satisfies the first law
\be
d\Phi = I_a  dE^a , \quad
I_a = \frac{\partial \Phi}{\partial E^a} .
\ee

As mentioned above, quasi-homogeneity is an important condition to obtain consistent results in black hole thermodynamics. We impose the same condition on the potential $\Phi(E^a)$, i.e., we assume that 
\be
\Phi(\lambda^{\beta_a} E^a) = \lambda^{\beta_\Phi} \Phi (E^a). 
\ee
This condition plays an important role in the formulation of GTD because the quasi-homogeneity coefficients enter the explicit form of the metrics of the equilibrium space as  \cite{quevedo2023unified,quevedo2024geometrothermodynamic} 
\be
g^{{I}} =  \sum_{a,b,c=1}^n \left( \beta_c E^c \frac{\partial\Phi}{\partial E^c} \right)  
\frac{\partial^2\Phi}{\partial E^a \partial E^b}  dE^ a d E^ b ,
\label{gdownI}
\ee
\be
g^{II} =   
 \sum_{a,b,c,d=1}^n \left( \beta_c E^c \frac{\partial\Phi}{\partial E^c} \right) 
 \eta_a^{\ d}
\frac{\partial^2\Phi}{\partial E^b \partial E^d} dE^ a d E^ b   ,
\label{gdownII}
\ee
\be
g^{{III}} = \sum_{a,b=1} ^n \left( \beta_a  E^a \frac{\partial\Phi}{\partial E^a}\right)
 \frac{\partial ^2 \Phi}{\partial E^a \partial E^b}
dE^a dE^b \ ,
\label{gdownIII}
\ee
where $\eta_a^{\ c}={\rm diag}(-1,1,\cdots,1)$.
Moreover, to further reduce the calculations, we can use 
 the quasi-homogeneous Euler identity and Gibbs-Duhem relationship \cite{quevedo2019quasi}, which can be written as  
\be
\sum_{a=1}^n \beta_a I_a E^a = \beta_\Phi \Phi ,
\quad
\sum_{a=1}^n \left[ (\beta_a-\beta_\Phi) I_a dE^a + \beta_a E^a dI_a\right] =0,
\label{identities}
\ee
respectively, and reduce to the usual homogeneous expressions for $\beta_a=\beta_\Phi = 1.$

For later use, we explicitly consider the case $n=2$ with the fundamental equation   $\Phi = \Phi(E^1,E^2)$. Then, the resulting line elements can be written as \cite{quevedo2024geometrothermodynamic} 
\bea
\label{gI2D} 
g^{I}& = & \Sigma \left[\Phi_{,11} (d E^1)^2 + 2 \Phi_{,12} dE^1 dE^2 + \Phi_{,22} (dE^2)^2\right]\, \\
g^{II} & = & \Sigma \left[-\Phi_{,11} (d E^1)^2  + \Phi_{,22} (dE^2)^2\right]\,,
\label{gII2D} \\
g^{III} & = & \beta_1 E^1 \Phi_{,1} \Phi_{,11} (dE^1)^2 + \Sigma \Phi_{,12} dE^1 dE^2
+ \beta_ 2 E^2 \Phi_{,2} \Phi_{,22} (dE^2)^2 \, ,
\label{gIII2D}
\eea
where $\phi_{,a} = \frac{\partial \phi}{\partial E^a}$ and $\Sigma  =  \beta_1 E^1\Phi_{,1}  + \beta_2 E^2\Phi_{,2} .
$  
If we apply the Euler identity, $\Sigma = \beta_\Phi  \Phi$, to analyze the corresponding curvature scalars, we obtain that the curvature singularities are determined by the conditions
\be
I: 
\Phi_{,11}\Phi_{,22} -(\Phi_{,12})^2
=0 \ , 
\label{condI}
\ee
\be
II: \Phi_{,11} \Phi_{,22} 
=0\ ,
\label{condII}
\ee
\be 
III: \Phi_{,12}= 0\ ,
\label{condIII}
\ee
respectively, which coincide with the stability conditions and determine the phase transition structure of a system with two thermodynamic degrees of freedom \cite{callen1991thermodynamics}. This explicitly proves  the connection between phase transitions and curvature singularities in GTD.


\section{The Schwarzschild black hole with R\'enyi entropy}
\label{sec:sch}

In this section, we investigate how the use of the non-extensive R\'enyi entropy affects the thermodynamic properties of the static Schwarzschild black hole. 
Recall that in this case the event horizon is located at the radius $r_h=2M$ and the area of the horizon is given by
 $A=4\pi r_h^2 = 16\pi  M^2$. Then, using the Bekenstein-Hawking entropy, this black hole can be considered as a system with only one thermodynamic degree of freedom ($n=1)$. Then, from  the first law, $dS_{BH} = \frac{1}{T}dM$, and the standard definition of the heat capacity $C_{BH}=T_{BH}\left(\frac{\partial S_{BH}}{\partial T}\right)$, we obtain that the main thermodynamic properties of the Schwarzschild black hole are given by  
\be
S_{BH} = 4 \pi M^2\ , \quad T_{BH} = \frac{1}{8\pi M} \ , \quad 
C_{BH}= -  \frac{ \left(\frac{\partial S_{BH}}{\partial M}\right)^ 2}
{\frac{\partial^2 S_{BH}}{\partial M^2}} = - 8 \pi M^2 \ .
\ee
A particular result of the approach with the Bekenstein-Hawking entropy is that this black hole is a completely unstable system, as follows from the negative value of the heat capacity. 

Considering now the Schwarzschild black hole as described by the R\'enyi entropy
\be
S  = \frac{1}{r} \ln(1+4\pi r M^2),
\label{schent}
\ee
we see that it corresponds to a system with two thermodynamic degrees of freedom $(n=2)$ since, as proved in Sec. \ref{sec:non}, the R\'enyi parameter $r$ is also a thermodynamic variable.

We will use GTD to investigate the properties of this black hole in an invariant way. As mentioned above, for the fundamental equation $\Phi = \Phi (E  ^a) $
we can use any thermodynamic potential. For simplicity, we will use the mass as the thermodynamic potential, which according to Eq.(\ref{schent}) can be expressed as 
\be
M(S ,r) = \frac{1}{2}\sqrt{ \frac{e^{rS }-1}{\pi r}} \ 
\label{mass}
\ee
and satisfies the first law
\be
dM = T dS  + R d r \ ,
\quad
T = \frac{\partial M }{\partial S }\ ,  \quad
R= \frac{\partial M }{\partial r}\ ,
\ee
where we have introduced the thermodynamic variable $R$ dual to the R\'enyi parameter $r$. 

\begin{figure}
    \centering
\includegraphics[scale=0.5]{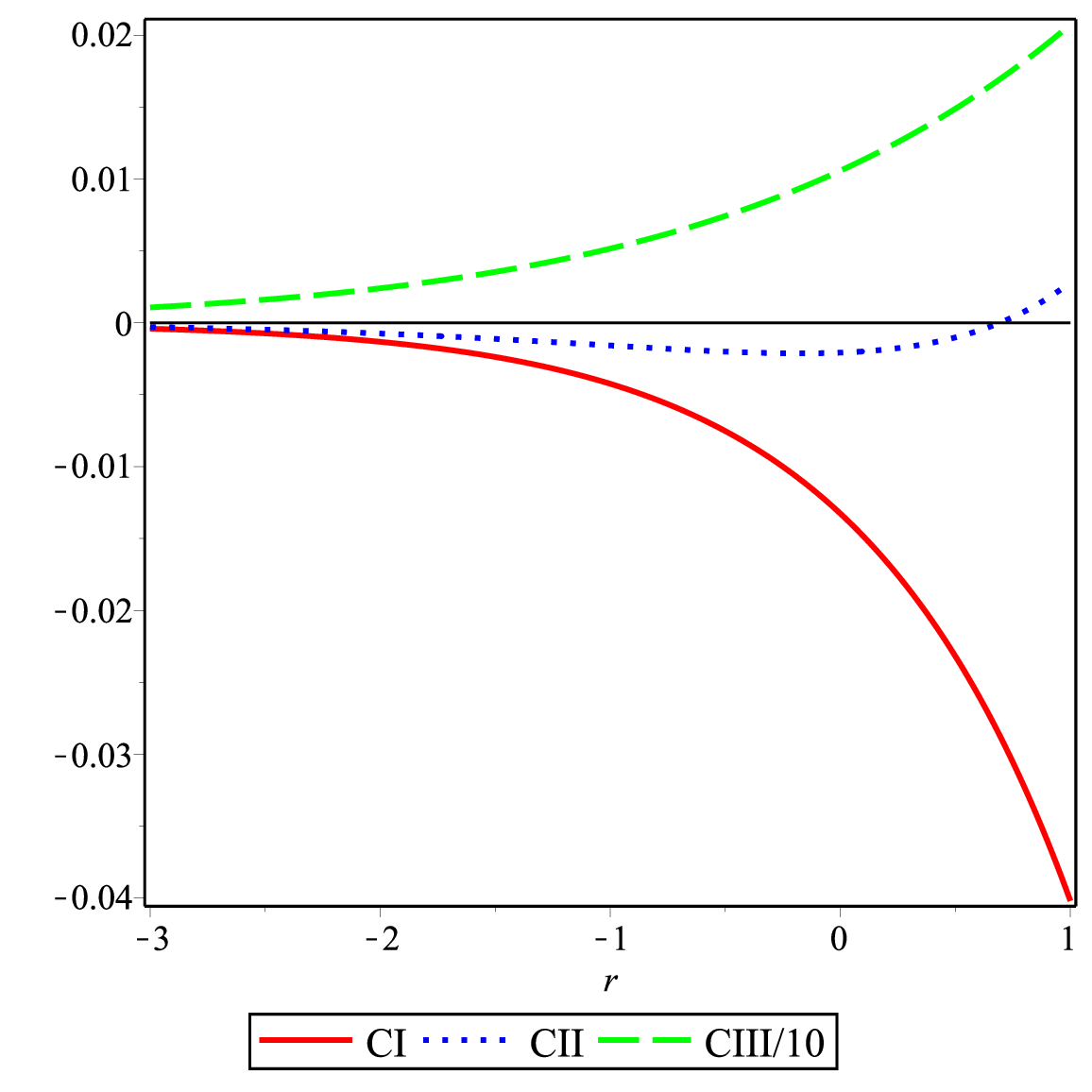}
    \caption{The singularity conditions for $S =1$ as functions of the non-extensivity parameter $r$. Only the curve $CII$ crosses the zero axis, where the curvature of the metric $g^{II}$ diverges. }
    \label{fig1}
\end{figure}

Applying the quasi-homogeneity condition to the mass function (\ref{mass}), we
obtain 
\be
\beta_{S } = -\beta_r,\quad 
\beta_M = - \frac{1}{2} \beta_r.
\ee
Then, the Euler identity (\ref{identities}) becomes
\be
S\frac{\partial M}{\partial S} - r \frac{\partial M}{\partial r}   = \frac{1}{2} M,
\ee
an expression that can be used to simplify the explicit form of the metrics. In fact, using the notation $\Phi=M$, $E^1=S $, and $E^2=r$, the metrics (\ref{gI2D})--(\ref{gIII2D}) can be expressed as
\be
g^I = -\frac{\beta_r}{32\pi (e^{rS }-1)}\left( F_1 dS ^2 + 2 F_2 dS \, d r + F_3 dr^2\right),
\ee
\be
g^{II} = -\frac{\beta_r}{32\pi (e^{rS }-1)} \left( - F_1 dS ^2 + F_3 dr^2\right),
\ee
\be
g^{III} = - \frac{\beta_r}{32\pi (e^{rS }-1)}\left[ \frac{S  e^{rS }}{4\pi M} F_1 dS ^2 + F_2  dS \, d r  -\frac{1+rS (e^{rS }-1)}{4\pi M^2}  F_3 dr^2\right],
\ee
where
\be
F_1 =  r e^{rS }(e^{rS }-2) ,
\ee
\be
F_2 =  \frac{e^{rS }}{r}[ e^{rS }(1+rS ) -2 r S  -1] ,
\ee
\be
F_3=\frac{1}{r^3 }
[e^{rS }(e^{rS }-2) (e^{rS }-rS +3) - rS  e^{rS } + 3].
\ee
Then, it is straightforward to compute the singularity conditions (\ref{condI})--(\ref{condIII}), which can be reduced to 
\be
CI: \quad  \frac{e^{rS }}{r^3 M^4}\left[ e^{rS }(1-2r S ) -4e^{rS }(1-rS ) + 3
\right]=0 ,
\ee
\be
CII: \quad \frac{e^{rS }(e^{rS }-2)}{r^4 M^6} \left[
e^{2rS }(r^2 S ^2 - 2 r S  + 3) - 2e^{rS }(r^2S ^2-rS  +3) + 3
\right]=0,
\ee
\be
CIII:\quad 
\frac{e^{rS }}{\lambda^2 M^3} \left[ e^{rS }(1+rS ) - 2r S  - 1)
\right]=0 .
\ee
In Fig. \ref{fig1}, we illustrate the behavior of the above conditions. We see that only the condition $CII$ can be satisfied for a particular value of the non-extensivity parameter $r$, indicating the presence of a curvature singularity, which corresponds to a 
second-order phase transition.

\section{Interpretation of the results}
\label{sec:res}

According to the results presented in the previous sections, to consistently consider the Schwarzschild black hole in the framework of the non-extensive R\'enyi entropy, it is necessary to modify its thermodynamic properties. In fact, the main thermodynamic variables are now the mass $M$, temperature $T$, and the dual $R$ to the non-extensivity parameter $r$, which are explicitly given by
\be
M = \frac{1}{2}\sqrt{\frac{e^{rS}-1}{\pi r}}\ ,
\quad 
T =  \frac{e^{rS}}{8\pi M} \ , \quad
R = 
\frac{e^{rS}(rS-1)+1}{8\pi r^2 M}\ ,
\ee
and are related through the Euler identity
\be
2TS -2r R = M .
\ee

\begin{figure}
    \centering
\includegraphics[scale=0.5]{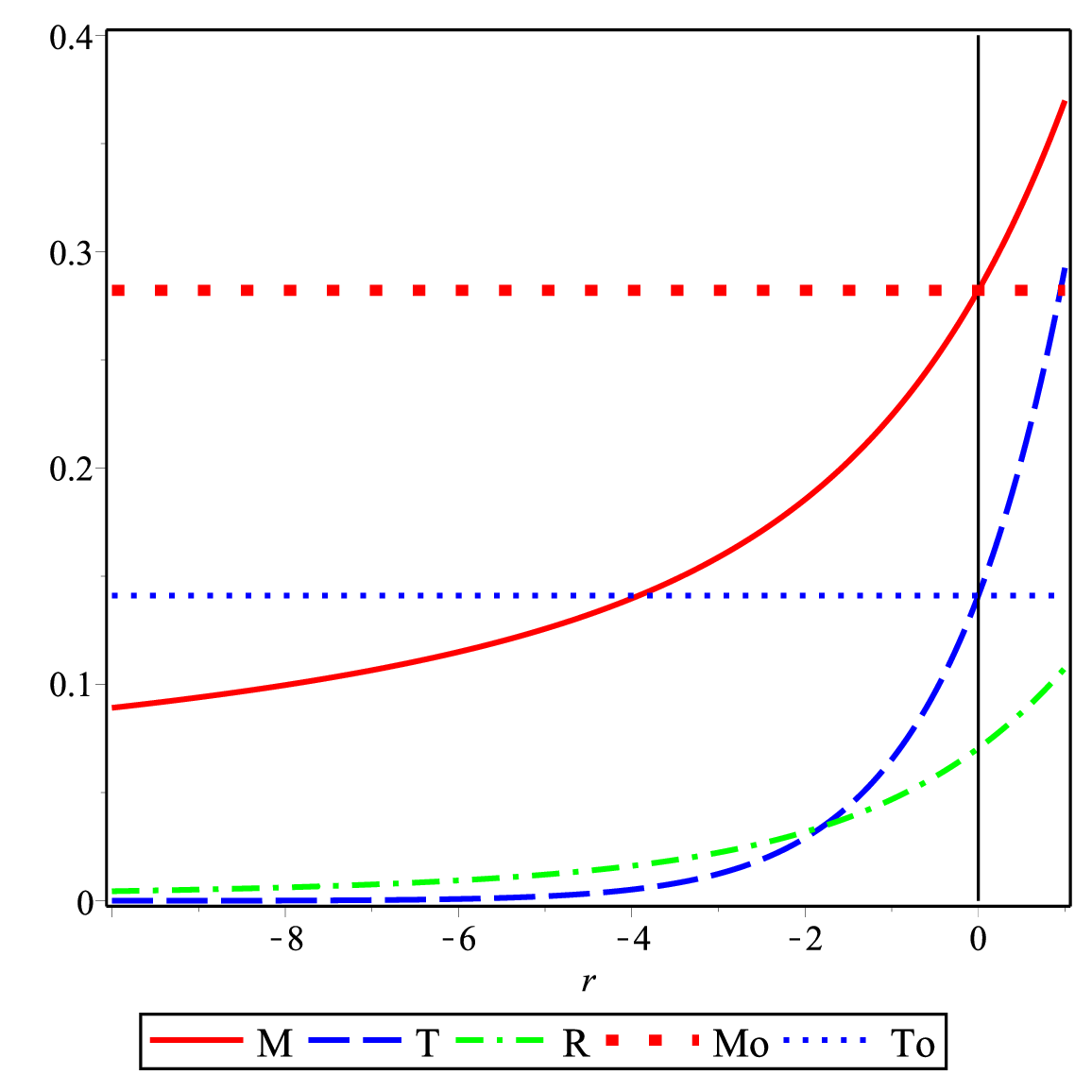}
    \caption{Mass $M$, temperature $T$, and dual non-extensivity $R$ of the Schwarzschild black hole with R\'enyi entropy as functions of the non-extensivity parameter $r$. The entropy is fixed as $S=1$.
    The doted lines represent the mass $M_0=\frac{1}{2} \sqrt{\frac{S}{\pi}}$ and temperature $T_0=\frac{1}{4\sqrt{\pi S}}$ for the Schwarzschild black hole with $r\rightarrow 0$. 
    }
    \label{fig2}
\end{figure}

The Hawking temperature $T_0= \frac{1}{8\pi M}$  results modified by the factor $e^{rS}$, and the variable $R$ is conceptually of a new nature.  The thermodynamic behavior of these variables is illustrated in Fig. \ref{fig2} in terms of the non-extensitivity parameter $r$. We can see that all variables are positive definite, reach their maximum value for $r\rightarrow 1$, and decrease as $r$ decreases. The values of the mass and temperature in the interval $r\in (0,1]$ are larger than the corresponding variables $M_0$ and $T_0$, with vanishing parameter $r$. 
Then, in the interval $r\in (-\infty,0)$, the mass and temperature are smaller than $M_0$ and $T_0$, respectively. 

\begin{figure}
    \centering
\includegraphics[scale=0.5]{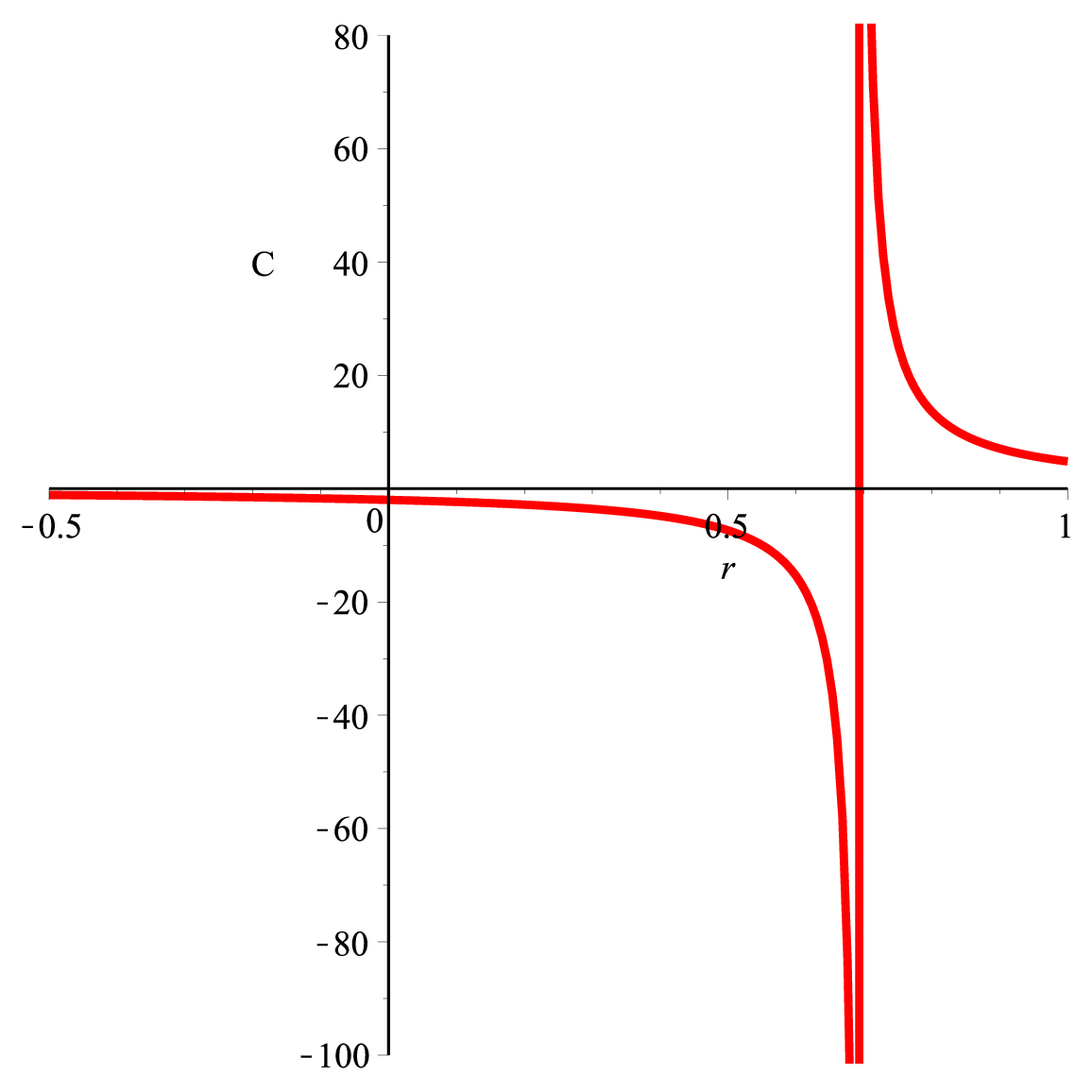}
    \caption{Heat capacity of the Schwarzschild black hole with  R\'enyi entropy for $S=1$ and different values of the non-extensivity parameter $r$. }
    \label{fig3}
\end{figure}

In the previous section, we showed that there is a second-order phase transition indicated by a curvature singularity associated with the metric $g^{II}$. According to the expression (\ref{condII}) for $CII$, the location of this curvature singularity corresponds, in general, to the condition $e^{rS}-2=0$. The phase transition in this case is associated with a change of stability of the black hole as can be shown by considering the heat capacity 
\be
C:= T \left(
\frac{\partial S}{\partial T}\right) = T \left(\frac{\partial ^2 M}{\partial S ^2}\right)^{-1} = \frac{2(e^{rS}-1)}{r(e^{rS}-2)}, 
\ee
which shows a divergence at $e^{rS} = 2.$ We depict the behavior of the heat capacity in Fig. \ref{fig3}, where the divergence shows a transition from a unstable state $(C<0)$ to a stable state $(C>0).$ This shows that non-extensivity drastically affects the thermodynamic properties of the Schwarzschild black hole, allowing the existence of a new phase  with stable states. The stable sector depends on the value of the non-extensive parameter and is located within the interval 
 $r\in \big(\frac{\ln  2}{S}, 1\big]$.

\section{Conclusions}
\label{sec:con}

In this work, we have investigated black hole thermodynamics in the framework of R\'enyi entropy, which is consistent with the classical and statistical approach to the laws of thermodynamics.
The main point is that R\'enyi entropy agrees with the non-extensive character of black hole thermodynamics.

In addition, we impose the property of quasi-homogeneity at the level of the fundamental equation to be in concordance with the non-extensivity of the entropy and the geometrothermodynamic approach to classical thermodynamics. As a result, we find that the non-extensivity parameter $r$, which is defined in the interval $r\in (-\infty,1]$, should be considered as an independent thermodynamic variable. 

{ Some concern could be raised regarding the interpretation of the R\'enyi parameter as a thermodynamic variable. In fact, this parameter was introduced originally at the mathematical level to generalize the Tsallis entropy. So, at first sight, it could be thought that a mathematical parameter cannot have any significance at the physical level. Nevertheless, the idea of interpreting  parameters of pure mathematical origin as physical parameters and thermodynamic variables is not new. In fact, one of the last applications of this idea has been realized in black hole thermodynamics, where coupling constants are interpreted as thermodynamic variables, giving rise to a completely unexpected field in which black hole configurations can be analyzed by using  concepts of classical chemistry \cite{mann2025black}.
In the case of the R\'enyi entropy, it has been argued  recently in \cite{nakarachinda2025thermodynamics} that to recover the Smarr formula and the first law of thermodynamics, the R\'enyi parameter can be interpreted as an effective pressure, i.e., a thermodynamic variable whose dual is identified as a thermodynamic volume.
The results of the present work corroborate the thermodynamic character of the R\'enyi parameter from a different perspective, namely, by imposing the physical condition of quasi-homogeneity and using the formalism of GTD.}

As a particular example of our approach, we investigate in detail the case of the Schwarzschild black hole in the framework of extended thermodynamics with the R\'enyi entropy and non-extensivity parameter as independent thermodynamic variables. 
We use the formalism of GTD to obtain invariant results that do not depend on the choice of thermodynamic representation or potential.

Analyzing the behavior of the GTD metrics, we established that the Schwarzschild black hole in extended thermodynamics can undergo a second-order phase transition, which changes the stability properties of the system, introducing a new phase of stable states.  
In general, we showed that the stable phase occurs within a specific interval of values of the non-extensivity parameter, namely, $r\in \big(\frac{\ln  2}{S}, 1\big]$.

{ Our results show that considering the R\'enyi parameter as a thermodynamic variable has a non-trivial physical significance, because it leads to important consequences affecting the stability properties and the phase transition structure of the static Schwarzschild black hole.  }

Here, for simplicity, we have analyzed only the static black hole in extended thermodynamics. Clearly, using the approach of GTD, it is possible to generalize our results to include stationary black holes. We plan to analyze this and other more general black hole configurations in future work.

\section*{Acknowledgments}

 This research was funded by the Vicerectoría de Investigaciones de la Universidad Militar Nueva Granada-Vigencia 2024, grant No. INV-CIAS 3918.
The work of HQ was supported by UNAM-DGAPA-PAPIIT, grant No. 108225, and Conahcyt, grant No. CBF-2025-I-243.

\bibliography{references}
\end{document}